\documentclass[structabstract]{aa}

\usepackage{graphicx}
\usepackage{txfonts}
\begin{document}

\title{A long term spectroscopic and photometric study of the old nova HR
Del}

\author{
M.Friedjung 
\inst{1}
\and M. Dennefeld 
\inst{1}\thanks{Visiting Astronomer, Observatoire de Haute-Provence, CNRS, France} 
\and 
I. Voloshina 
\inst{2}\thanks{Visiting Astronomer, Crimean station of the Sternberg Astronomical Observatory }
}

\offprints{M. Friedjung}
\mail{fried@iap.fr}

\institute{Institut d'Astrophysique de Paris -UMR 7095, CNRS/Universit\'e 
Pierre et
Marie Curie, 98 bis Boulevard Arago, 75014 Paris, France \\ 
\and
Sternberg Astronomical Institute, 119992 Moscow, Russia}
\date{Received December 22, 2009;  Accepted June 17, 2010}

\titlerunning{The old nova HR Del}
\authorrunning{Friedjung et al.}

\abstract
{The Nova HR Del, discovered in 1967,  was found to be exceptionally
bright in the optical and UV during
the whole lifetime of the IUE satellite (ending in 1996) and appears to
be still
extremely luminous today. The reason for
this continuing activity is not clear; continuing weak thermonuclear burning
might be involved.}
{We  therefore need to better characterise and understand the processes at play in
HR Del.}
{HR Del was thus monitored over several years, both in broad band photometry
and spectroscopically in the H$\alpha$ spectral region.}
{The profile of the H$\alpha$ line shows two components: a narrow, central
component and broader wings. The former  is most easily understood as
being due to an accretion disk, whose geometry might lead to it partly
occulting itself. That component shows something like an S wave with an
orbital
phase dependance, suggesting that it could be due to a spot bright in
H$\alpha$. The wide component must come from another region, with a
probably non-negligible contribution from the material ejected during the 1967
outburst.
Non-orbital variations of the H$\alpha$ equivalent width
were found both on long and short time scales. Similar variations were
found in the photometry, showing a 
component with a clear dependence on the orbital phase, but no obvious relation with the H$\alpha$ variations.}
{The orbital part of the photometric variations can be explained by irradiation
of the companion, while the properties of H$\alpha$ are explicable by the
presence of an accretion disk and a spot bright in H$\alpha$.}


\maketitle
\section{Introduction}

HR Delphini (or Nova Del 1967) was a bright classical nova with however
unusual properties.  It was discovered on July 8.9 1967 by Alcock (see Candy, \cite{Cand67}),
when it had brightened to a magnitude near 5.5 from a pre-outburst magnitude of
about 12 at the beginning of June. After remaining 5 months near this
magnitude, it brightened again to a short peak at m$_v$ of 3.5 in December 1967,
which probably corresponds to what is considered to be the maximum for classical novae. 
This was followed by an extremely slow irregular decline, as can be seen for instance
in Sanyal (\cite{Sany74}). During the stage before the brightest peak, the velocity of the main
absorption component continuously decreased, unlike the usually observed behaviour of
 velocities in   
 classical novae after maximum: this reinforces the idea that the short peak in December 1967 
was indeed the "real" maximum. The unusual behaviour
of this nova in such an exceptionnally long pre-maximum stage might, according to
Friedjung (\cite{Fried92}), be explainable by the presence of a wind, 
 optically thin in the continuum, and with a decreasing velocity during that
pre-maximum stage, unlike the velocities of the optically thick winds normally seen in
classical novae after maximum. \object{HR Del} may, indeed, only have satisfied
marginally the conditions needed for thermonuclear runaway. This could be due
to the very low mass  (0.55-0.75 M$_{\odot}$) of the white dwarf component of the binary 
(Selvelli and Friedjung, \cite{Selv03}). A few other
classical novae such as \object{V723 Cas} (1995) may have a similar nature.

In addition, this nova showed signs of unusual activity both before and long
after its outburst (Selvelli and Friedjung \cite{Selv03}). This can be seen by
its exceptional luminosity, compared with that of other old novae both from IUE
observations in the UV and from its optical apparent brightness. The time
variable P Cygni profile of the CIV resonance doublet also indicated a very
high wind velocity of the post-nova of the order of 5000 km s$^{-1}$. The
deduced values of distance of 850 pc and $E(B-V) = 0.16$ led to a UV luminosity
of 56 $L_{\odot}$ and a visual absolute magnitude of $M_v$ = 1.85. The latter
value has to be corrected for the inclination of the system: assuming that the
radiation came from a gravitationally heated accretion disk (then having an
accretion rate of $dM/dt$ = 1.4 10$^{-7}$M$_{\odot}$ $yr^{-1}$ onto a white dwarf with
a radius of 0.0125$R_{\odot}$), the correction leads to an absolute magnitude of $M_v$
= 2.30. Such observations suggested unusual activity: one possible explanation
was continuing very weak thermonuclear burning, with the radiation of the white dwarf being
reprocessed by an accretion disk (and the companion star), 
a situation theoretically possible when the white dwarf component's
mass is low  (Sion and Starrfield \cite{Sion94}). In this connection we can
note that V723 Cas was  still an active X-ray source more than 12 years after
the outburst (Ness et al. \cite{Ness08}), but in the case of HR Del, only weak X-ray  
emission  was reported at early times when instruments were much less sensitive
than now (Hutchings \cite{Hutch80}, with 6$10^{31}$ ergs s$^{-1}$ after correction for our adopted distance), 
and nothing more recently.

It is  possible to obtain a better clue to the luminosity by estimating the ionisation
of the nebular remnant by the central source. The remnant of \object{HR Del}
has been studied in a number of papers, e.g. Harman $\&$ O'Brien (\cite{Harm03}). 
Very recently, Moraes and Diaz (\cite{Mora09}) studied the clumpy structure
of the nebula. Detailed photoionisation modelling suggest the presence of a
disk shaped ionising source, with a
luminosity of $10^{36}$ ergs s$^{-1}$ and a temperature of 65,000K. The two
sides of a disk, emitting like a black body at such a temperature, would then
have a radius of 1.3 $10^{10}$ cm or 0.2 $R_{\odot}$. Such a relatively low
temperature might explain why the X-ray emission is weak . A total shell mass
of 9 $10^{-4} M_{\odot}$ was also found by these authors, at least half being
contained in the neutral clumps.

In the light of these results, time variations of HR Del need to be followed
by ground based observations, as they could reveal some signs of exceptional
activity. It was with this in mind that we observed the region of H$\alpha$
with a high spectral resolution over several years, as this line  could
give particularly significant results. Parallel broad band photometric
monitoring was also performed. Some preliminary results of this observing campaign 
 have already been presented by Friedjung et al. 
(\cite{Fried05}).

\section{Observations}

{\bf Photometry} \par \bigskip

\begin{figure}
\centering
\includegraphics[bb=15 15 310 250, width=7.5cm, angle=0, clip]{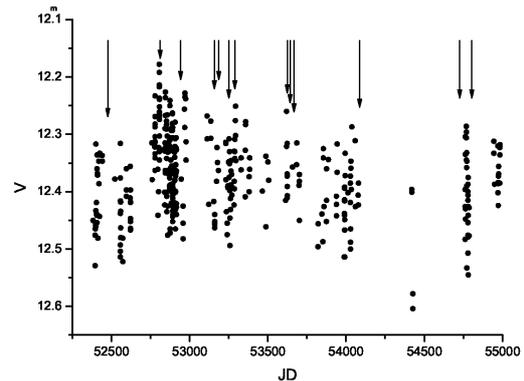} 
\caption{ Overall behaviour of HR Del in the $V$ band during 2002-2008.
Photoelectric V magnitudes are shown 
versus Julian Day. The arrows indicate the 
epochs of the spectroscopic observations.  } 
\label{Overall}
\end{figure}

Photometric observations of HR Del were made in order to detect possible
long term variations, and also to verify  the orbital phases for the
spectroscopic observations. They were carried out with two different
instruments:  firstly, the single channel  photoelectric photometer on
the 60-cm telescope of the Sternberg Astronomical Institute in Crimea was
used from the  beginning in 2002 till 2009, with continuously changing
Johnson UBV filters and a 10 s  integration in each filter. The usual 
method of differential measurements was applied. A local standard was
used permanently during observations, and 
calibrated against the well-known standard star SAO 106418  
(magnitudes  $V=8^m.59, B=8^m.62, U=8^m.55$). 
More than 1600  measurements  of HR Del were obtained in 3 bands during the
total time  of our observations, with usually 2-3 measurements per night.
These data were used to determine the long-term behaviour of HR Del, which is 
shown in Fig.~\ref{Overall}. They  are accurate to $1\%$ in the $V$ and the
 $B$ bands and $2\%$ in $U$ band. 

\begin{figure}
\centering
\includegraphics[bb=15 15 262 316, width=7.5cm, angle=0, clip]{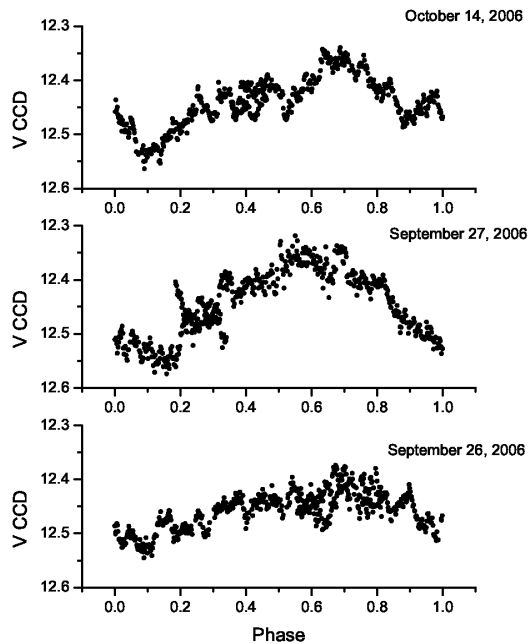} 
\caption{  Examples of CCD observations for different dates: CCD-V magnitudes versus phase. } 
\label{CCD}
\end{figure}

Secondly, CCD observations of HR Del were obtained  with the goal of studying
variability on time scales from minutes to days. Several runs of a few nights
were made in August and October 2005, during the period of July-October 2006,
then during  October-November 2008, and finally during three nights in October
2009. Most of these  observations were carried out in the $V$ band with the
60 cm telescope of the Crimean station of the Sternberg Astronomical Institute,
using an Apogee 47  CCD detector with 1024 x 1024, 13 $\mu$m pixels, and a field
of view of 6 x 6 arcmin. During 2008, in addition to the above instrument used
both in the V and R bands, the 38cm telescope of the Crimean Astrophysical
Observatory was also used from November 10 to 13th, but only in the R band, 
with a CCD-ST7 camera (460 x 700, 9$\mu$m pixels). In 2009, a Pictor
MEADE-416 camera was also used on the 50 cm Sternberg telescope in the V band. 
All these data were obtained in a 2x2 binning mode  to reduce the read-out time.  
The duration of individual observational series varied from  2.5 up to 4.5
hours. A star close to HR Del in the field was always used as a local 
standard (a few suitable  stars were used,  among them the one used for the
photoelectric observations), 
and  repeatedly calibrated using  well-known comparison stars during a few nights with perfect  weather. 
Another two stars with coordinates $\alpha=20^h42^m18^s$ and $\delta= +19^{\circ}08'30"$ $V=13^m.660, B=14^m.70, U=15^m.70$ and $\alpha=20^h42^m33^s$ and $\delta= +19^{\circ}06'30"$ $V=12^m.929, B=14^m.05, U=15^m.03$ were used as control stars. 
More then  4200 measurements of HR Del were obtained this way in the $V$ band during 2005-2008, covering more than 30 nights, while about 700 datapoints were obtained in R over 9 nights. 
The CCD measurement errors do not exceed $3\div4$\% in the V band, and slightly less for the R band. 
The details of the observations are presented in Table \ref{ObsPhot}. 
Examples of  daily light curves are presented in  Fig.~\ref{CCD}.   
A more detailed description  of these observations will be  presented in a companion paper (Voloshina et al., in preparation). \\

\bigskip

{\bf Spectroscopy} 
\par \bigskip

The spectroscopic monitoring was done  at the Observatoire de Haute-Provence   (OHP) in
France, using the 1.52m telescope, equipped with the Aur\'elie high-resolution
spectrograph (Gillet et al. \cite{Gill94}). Light passing through a 3" circular entrance
aperture then encounters a modified Bowen-Walraven image slicer (Walraven
\cite{Walr72}), whose 5 slices of 0"6 each are
imaged onto a  2048x1024 EEV CCD with 13.5 $\mu$m pixels. To gain in sensitivity,
all the corresponding lines of the CCD were summed before read-out, 
producing a single line spectrum as output. However 
as a consequence, the possibility of identifying and cleaning cosmic rays 
hits before final reduction disappears. The spectra were reduced following 
standard procedures: bias and flat-field corrections, and wavelength 
calibration using a thorium-argon internal lamp. In view of the small 
spectral range covered and used, no spectral response correction was 
performed. \\

\begin{figure}
\centering
 \includegraphics[bb=40 95 560 800, width=6.5cm, angle=-90, clip]{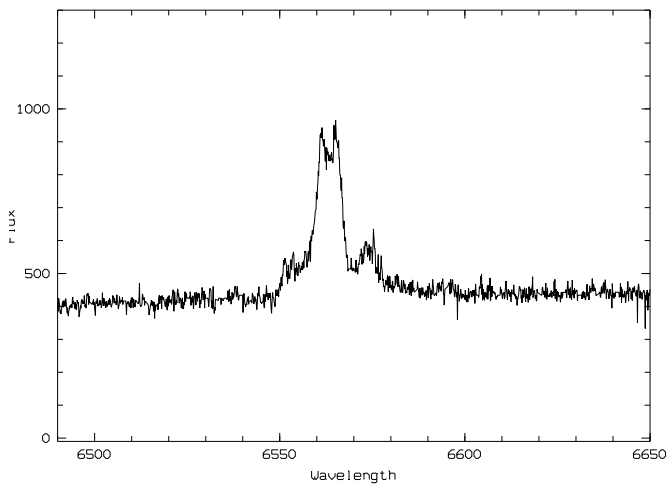}
\caption{Example of a high dispersion  spectrum, obtained on November 1th, 2003
at the OHP. 
The core, and the low intensity wings of  H$\alpha$ are easely distinguishable.}
\label{high}
\end{figure}

Many of the observations (but not all)  were done in service mode, and used 
a 1200 l/mm grating giving a dispersion of 7.6$\AA$ per mm, and a spectral resolution 
 of 0.32 $\AA$ (0.10 per pixel). Typical individual exposure times are 30 minutes, 
but the sequencing  of observations, and phase coverage is irregular, a 
tribute to be payed to the service mode operation. The details of 
observations are given in Table \ref{ObsSpec}. 
An exemple of a high-resolution spectrum is given in Fig.~\ref{high}. \\  

\begin{figure}
\centering 
 \includegraphics[bb=40 95 560 800, width=6.5cm, angle=-90, clip]{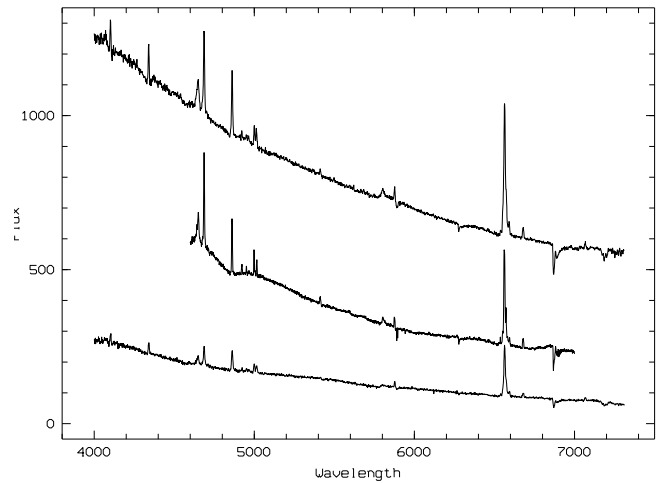}
\caption{Low dispersion  spectra. Lower part: OHP, Dec. 2007; Middle: 
INT, June 2008. Upper part: OHP, Dec. 2008. Small  offsets applied between spectra (see text). 
The units in the ordinates are 10$^{-16}$ ergs/cm$^{2}$/sec/$\AA$.}
\label{low}
\end{figure}

A few, low dispersion spectra were obtained towards the end of the campaign,
to check the general aspect of the object. We used the OHP 1.93m telescope 
in December 2007, with the Carelec spectrograph and a 300 l/mm grating 
giving 1.8 $\AA$ per pixel and a 6.5 $\AA$ resolution with the 2" slit used. 
The same setting was used again in December 2008. 
In June 2008, we used the IDS at the INT 2.5m telescope in La Palma: there, a  
630 l/mm grating gave 0.9 $\AA$ per pixel and 2.7 $\AA$ resolution with 
a 1"5 slit. Those spectra were reduced with standard procedures, including 
the use of standard star observations to correct for the wavelength dependant spectral response. In 
the case of the INT however, only the central part of the spectrum can be 
exploited due to strong vignetting. Summation has been made over the full H$\alpha$ extent along the slit and 
the spectra are displayed in 
Fig.~\ref{low}. \\

\section{Results and analysis}  
\bigskip

{\bf A) Analysis of the photometry} \par \bigskip

As can be seen from Fig.~\ref{CCD}, daily light curves of HR Del show night to
night variations and also aperiodic stochastic variations, that is so called flickering.

\begin{figure}
\centering
\includegraphics[width=\linewidth, angle=0]{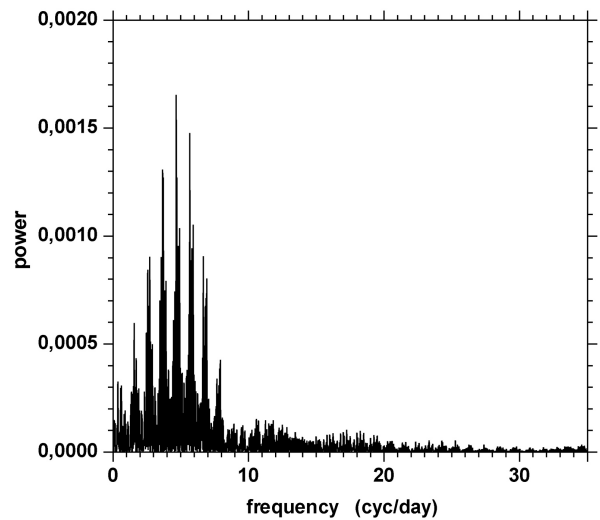} 
\caption{  Power spectrum of the 2008 CCD observations. }
\label{R_Power} 
\end{figure}

\begin{figure}
\centering
\includegraphics[bb=15 15 320 245, width=7.5cm, angle=0,
clip]{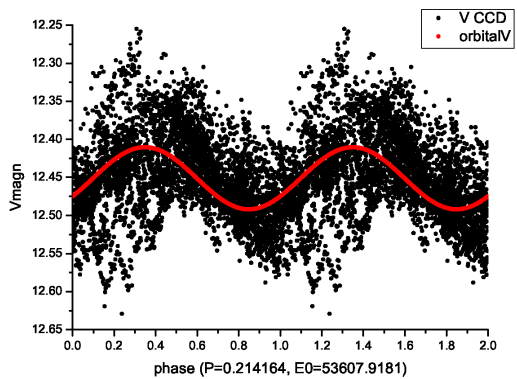} 
\caption{ Light curve of HR Del versus photometric phase, in the $V$ band, from observations
obtained  during 2005-2006. The averaged orbital sine wave is superposed onto the observed data.  } 
\label{li-curva}
\end{figure}

The CCD data, because they contain long time series of observations at similar
epochs, were used to search for the orbital period of the light curve (the
photoelectric data, although covering a larger total timespan, are too scarcely
sampled for that purpose). The analysis was made preferentially in the V band,
where more data are available, but the R band data were used also to check
for possible wavelength-dependant effects. They  were analysed using the
"Period04" PC code from Lenz \& Breger (\cite{Lenz04}) and also a programme
based on the Deming (\cite{Deming75}) method, in the frequency range
0-50 $d^{-1}$. The resulting periodogramm is displayed in Fig.~\ref{R_Power}:
the most prominent peak is for a period of 0.214164 days, corresponding,
within the uncertainties, to the period obtained by K\"urster and Barwig
from their spectroscopic analysis (0.214165), the other  peaks  being daily
aliases. Our observations folded with this period are shown in
Fig.~\ref{li-curva}. The scatter which remains in this figure is  due to
the presence  of night to night variations and possibly also flickering.   
Attempts to fold data with periods corresponding  to other peaks in the
periodogram lead to unacceptable fits, thus confirming the real period.  
If we approximate this light curve with a sine function (thick, red line),
the resulting semi-amplitude is  A = 0.043 $ \pm $ 0.002. 

No other short term variations were found in 
these observations besides the orbital ones. 
To better detect long timescale variations, such as those found by Bianchini (\cite{Bian87})
in some old novae, our data were corrected 
for orbital variations. 
There is a suspicion of
long-term variations with  a  possible period of around 1500 days (that is about 4 years), 
already apparent in Fig.~\ref{Overall},  
but several more years of  observations would be needed to confirm this. \\

\bigskip

{\bf B) Analysis of the H$\alpha$ spectral region. } \par \bigskip

The H$\alpha$ profiles  show a well defined central component, with
wider wings of lower intensity (Fig.~\ref{high}), 
suggesting formation in different regions. In view of this, 
measurements were made both of the total equivalent width and of that of the
line centre only.

\begin{figure}
\centering
 \includegraphics[bb=30 40 580 750, width=6.5cm, angle=-90, clip]{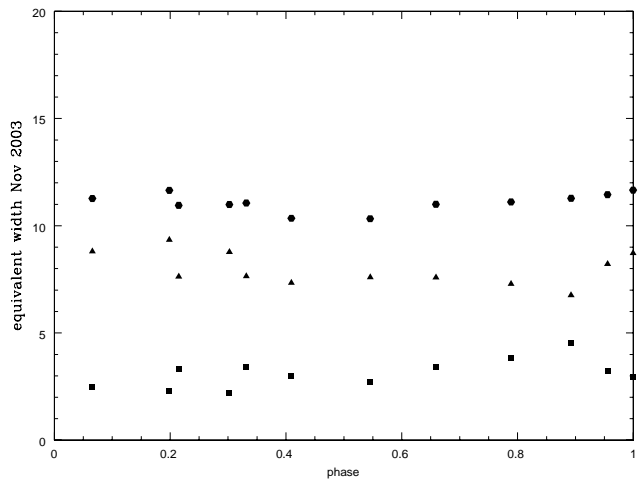} 
\caption{Equivalent width variation with orbital phase, November 2003. Filled circles (above) represent 
the total equivalent width; triangles (middle) the central component; and the squares (lower) the wings only. }
\label{nov03}
\end{figure}

\begin{figure}
\centering
 \includegraphics[bb=30 40 575 741, width=6.5cm, angle=-90, clip]{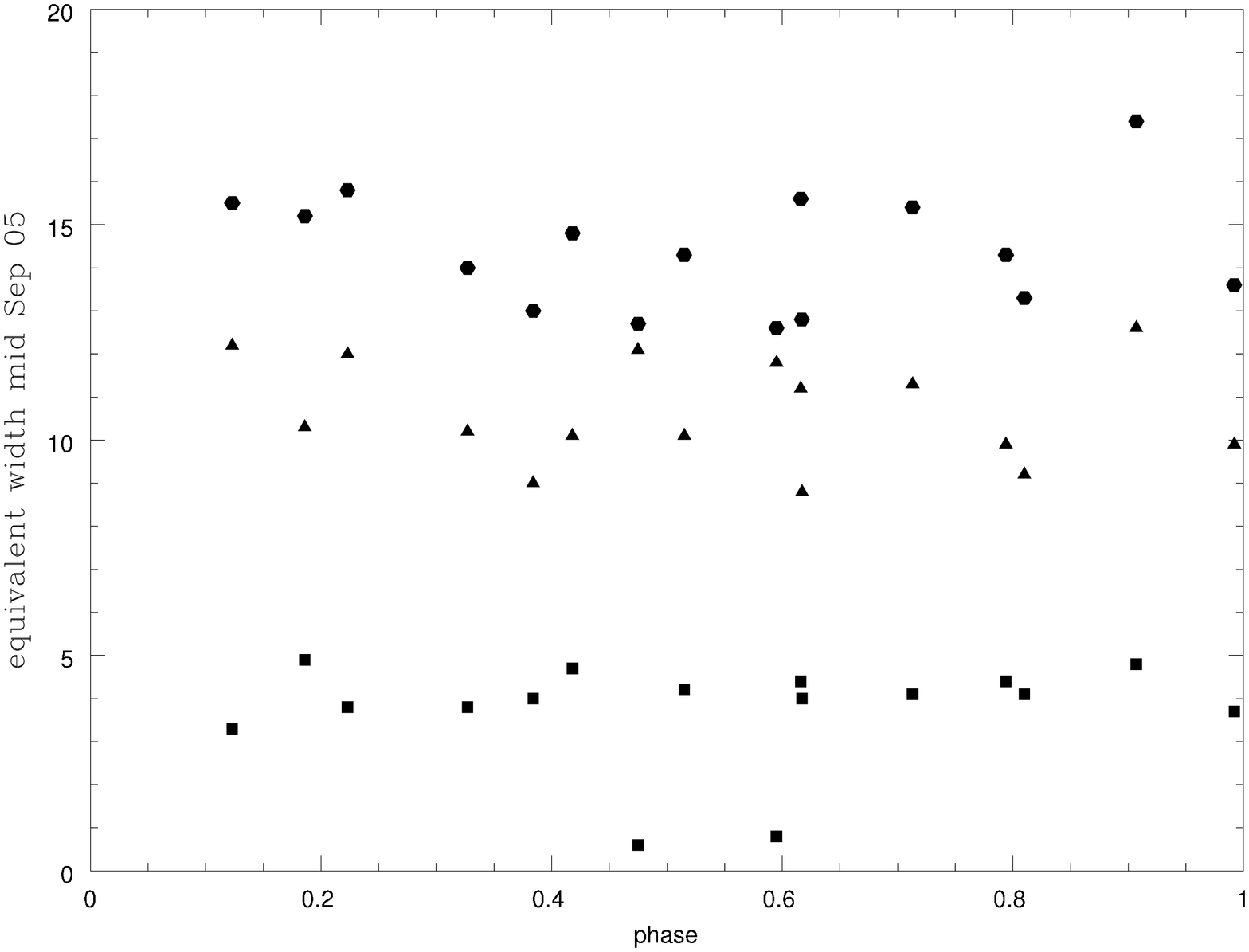} 
\caption{Equivalent width variation in mid September 2005. Symbols as above.}
\label{earlysep05}
\end{figure}

The time of each spectrum (at mid-integration), the
corresponding orbital phase according to the ephemeris of K\"urster and Barwig
(\cite{Kurs88}) and the measured  equivalent
widths of H$\alpha$ (both total, and central component only) 
are given in Table 1. Examples  of the variation of the equivalent
width of the different H$\alpha$ line components with Julian date and with K\"urster and
Barwig's orbital phase are shown for two epochs in Fig.~\ref{nov03} and
Fig.~\ref{earlysep05}, over a time interval not lasting more than a few days. 
Clear differences are seen between these two examples: on one hand, the average value 
of the  H$\alpha$ equivalent width is much lower at the first epoch; on the other hand, 
the scatter of the variations with phases is more important during the second epoch. Such changes are 
seen over the whole duration of our observing campaign. \\ 
One concern is a possible, time variable, contamination by emission from the ejected
nebula in the H$\alpha$ line due to the central region (plus a much smaller
contribution from the fainter, adjacent \ion{[N}{ii]}   emission).  
To estimate the possible contribution from the nebula, several tests have been made.
A detailed map of the nebular emission was obtained by Harman \& O'Brien
(\cite{Harm03}), combining  HST images with ground-based, high-resolution
spectroscopy. They showed that the nebular emission has an elliptical shape, with
major and minor axes of roughly 8.5 x 6.3 arcseconds, with a large 
H$\alpha$ contribution coming from a bright rim: these dimensions are much
larger than the 3" entrance aperture, centered on the star, used for our high
dispersion spectroscopy, so that the contribution from the nebula should not be
significant in the central  H$\alpha$ component. \\
The structure of the nebula, as seen in the HST images of Harman \& O'Brien
(\cite{Harm03}), is however clumpy, and several bright knots, located inside
the ellipsoid, had been identified. To measure their possible contribution to 
the H$\alpha$ flux from the central object (integrated over a circular aperture,
like the 3" used), it would be sufficient to measure the intensity of those blobs 
relative to the one of the central star. Unfortunately, the central object is
saturated in Harman's data, so we cannot estimate this ratio from there. We used
instead several ground-based images of ours, obviously of lesser spatial
resolution, but where some of the brightest knots can be identified, and found
that the intensity ratio between the 
brightest knots and the central star is typically 1 to 50 in H$\alpha$. It
would therefore require several knots within the aperture to make a 10\%
contribution, and only few are seen in the whole ellipsoid in the HST map. \\
We used also the spatial resolution along the slit in our low-dispersion
spectra to estimate the change in  H$\alpha$ equivalent width when including
more and more outer regions, and found also that 10\% was a conservative
upper-limit for a variable contribution from the nebula (see below the
discussion of the low dispersion spectroscopy). Finally, we note that the
measured velocities do not correspond either, at least for the central component, 
to what is expected from the nebula (see discussion later). \\
In view of all this, we shall consider  H$\alpha$  emission line equivalent
width variations of more than 10\% as intrinsic, as can be seen in the figures. 
The smallness of the contamination is understandable, following the expansion
and fading of the nebula, as  even our earliest observations were already about
35 years after the outburst of the nova. \\
\bigskip

As the line wings are always faint, the value of the total H$\alpha$ equivalent
width is dominated by the value of the narrow central component, so the variations
of these two components are similar.  
However it is not clear to what extent any of the orbital phase variations
are significant. In some cases, like November 2003 (Fig.~\ref{nov03}) or October
2005, the variations are small: for instance, on November 1-2 2003,  the  ratio
between the largest and smallest value of equivalent width is only 1.13. This can
give a limit to deviations from circular symmetry at
that time. In other cases, e.g. September 2005 (Fig.~\ref{earlysep05}), or August
2008, the variations, and the scatter are much larger. As in those 
longer series, more observing points are available, these larger variations are
probably indicative of rapid variations from night to night. \\ 

\begin{figure}
\centering
 \includegraphics[bb=30 40 580 750, width=6.5cm, angle = -90, clip ]{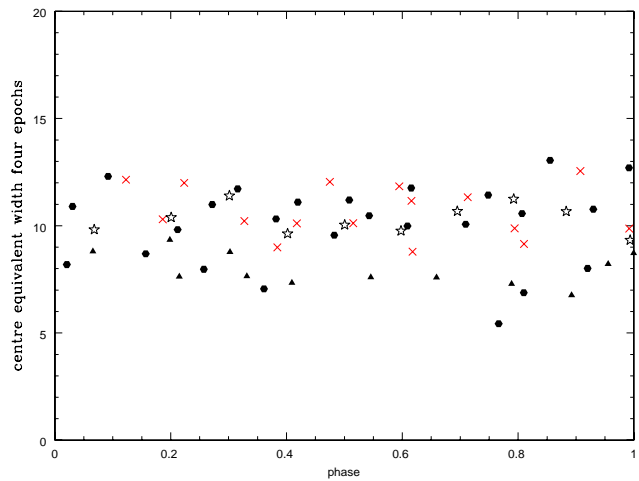} 
\caption{Equivalent width variation of the central component with phase, for four different observing periods. Triangles are for November 2003, crosses for September 2005, open stars for October 2005 and filled polygons for August 2008. }
\label{TotalFour}
\end{figure}

There are however significant differences in average equivalent widthes from
one epoch to another, as is illustrated in Fig.~\ref{TotalFour} for the central
component:  lower values are found for the 2003 epoch, with comparatively stable
values over different phases, while higher values are found in September 2005, and
a much larger change with phase in 2008. 

Quite large, longer time scale variations of the total and line centre
equivalent widths are seen (e.g. Fig.~\ref{centreall} for the central component).
 The values are clearly smaller in
2003 and 2004 than earlier or later. Such variations would be hard to explain
by contamination from varying contributions  of flux from the
ejected nebula. For the wings, no clear systematic variation with time of their 
equivalent width can be established, as the scatter is large; but changes in their
profiles and/or terminal velocity are definitely seen. 

\begin{figure}
\centering
 \includegraphics[ width=\linewidth ]{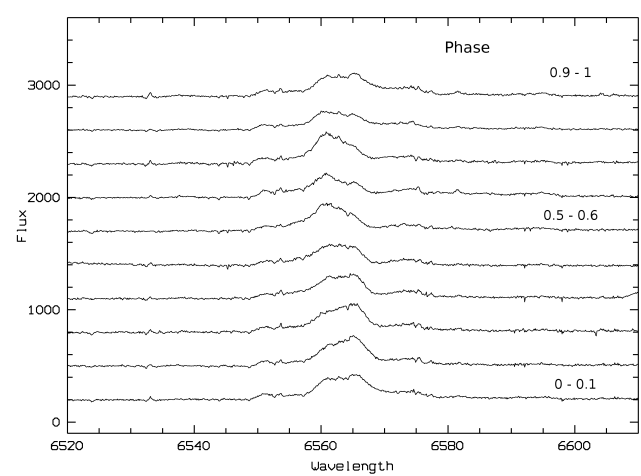} 
\caption{Series of profiles obtained during 2002-2004, averaged in ten bins of 0.1 in phases. The phases are going 
from 0 at the bottom (bin 0 to 0.1) to 1 at the top (bin 0.9 to 1.}
\label{SwaveB}
\end{figure}

\begin{figure}
\centering
 \includegraphics[ width=\linewidth ]{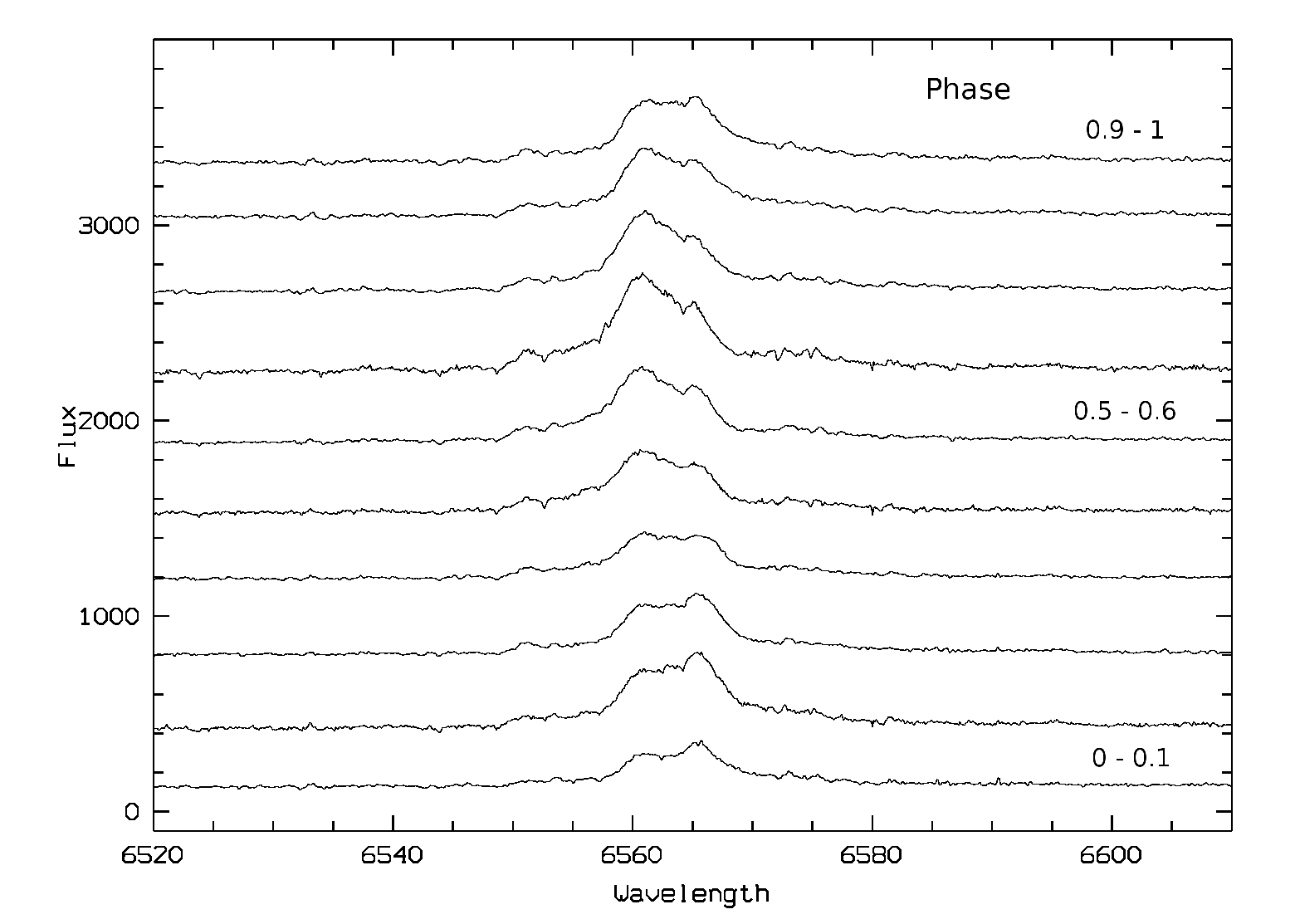} 
\caption{Series of  averaged profiles obtained in 2005 (bining and phases as above).}
\label{SwaveC}
\end{figure}

\begin{figure}
\centering
 \includegraphics[ width=\linewidth ]{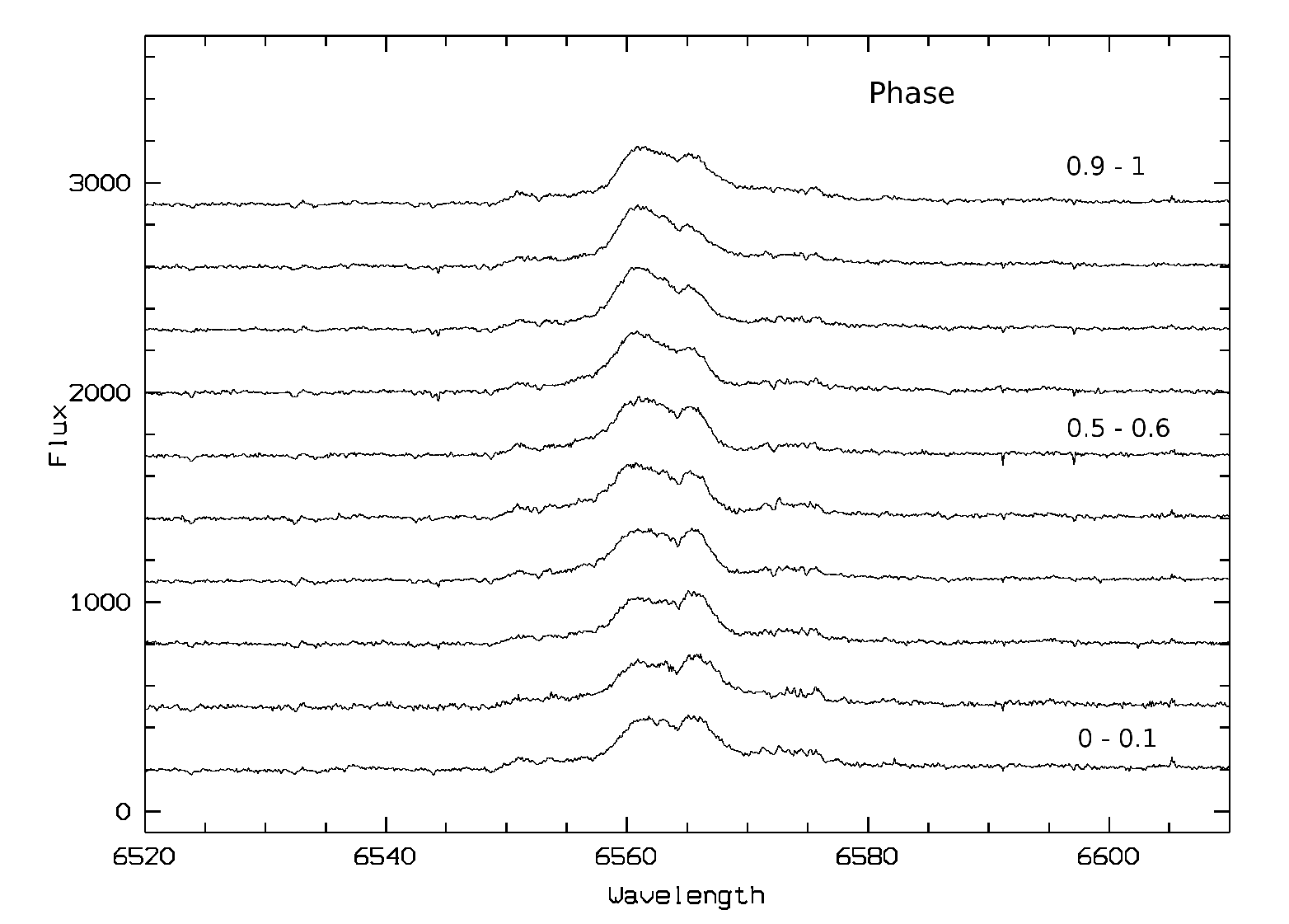} 
\caption{Series of averaged profiles obtained in 2008 (bining and phases as above).}
\label{Swave}
\end{figure}

\begin{figure}
\centering
 \includegraphics[bb=30 40 580 750, width=6.5cm, angle=-90, clip]{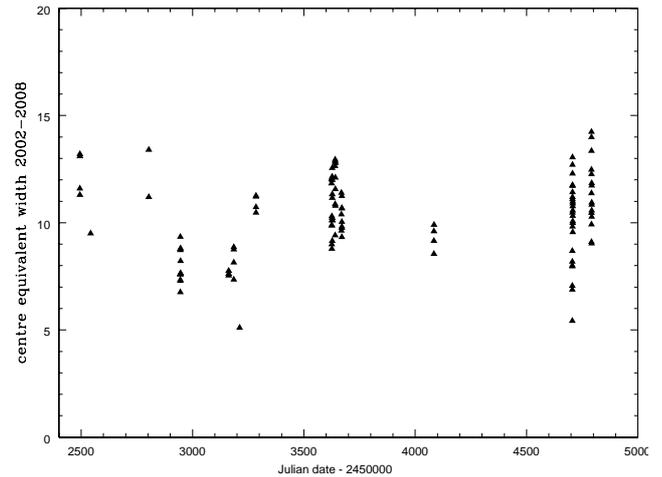}
 \caption{Long timescale equivalent width variation of the central component only.}
\label{centreall}
\end{figure}

The central component of the H$\alpha$ profile is assymmetric and varies with
orbital phase, suggesting something like the well known S wave of cataclysmic
variables. Comparison of profiles from successive spectra, obtained  at some
observational epochs when fairly long time series of observations could be
made, reveals this effect.  
Three examples are displayed, from the beginning, middle and  end of the campaign,
in Fig.~\ref{SwaveB}, \ref{SwaveC} and ~\ref{Swave} respectively, where series
of spectra from the same epoch have been bined over 0.1 intervals of phases. 
They show the permanency, and the stability of the phenomenon. The detailed profiles
may not look exactly the same, for a given bin, at each of the three epochs displayed,
but these small differences are due to the averaging process: although 
the total number of observed spectra was quite large, the number of spectra available
for a given bin at a given epoch is  
small, varying between 1 and 5, and furthermore the phase distribution  within a
given bin is quite inhomogeneous, so that the "average" aspect  can be strongly
influenced by the quality or phase of a single spectrum in the average. 
Nevertheless it is clear that the main features, like the phase of the maximum blue
or red peak in the central component  is quite clear, and reproducible.  
This confirms also that the spectroscopically determined period of 
K\"urster and Barwig (\cite{Kurs88}) is the most appropriate for our analysis. \\    

\bigskip

{\bf C) Low dispersion spectroscopy} \par \bigskip

The three spectra shown in Fig.~\ref{low} are rather similar: the  lower 
one was obtained at the OHP on Dec. 10, 2007 through thin clouds 
(exposure of  10mn), while the upper one  was obtained at the OHP on Dec. 7, 
2008 in good conditions (exposure of 5mn); the middle spectrum is from
the INT on June 25, 2008 (exposure of 200 sec). Note that the extremities
of this INT spectrum (bluewards of $\sim$~5000$\AA$ and redwards of
$\sim$~7000$\AA$) are not usable for determining the fluxes because of the
onset of a strong vignetting in the camera, which is difficult to correct
for in the reduction (the  bump seen in the continuum  around 5000 $\AA$
is also artificial). All three 
spectra are calibrated, and the differences in ordinates mainly reflect the 
difference in atmospheric transparency at the moment of the observations (except 
that the upper spectrum has been offset upwards by 300 units for clarity).  
The emission features detected comprise the Balmer lines of hydrogen
(H$\alpha$, H$\beta$, H$\gamma$, H$\delta$), lines of neutral helium (7065,
6678, 5876, 5015, 4921 $\AA$), the HeII lines at 5411 and 4686 $\AA$ and the
broad CIII/NIII $\lambda\lambda$ 4640 and CIV $\lambda\lambda$ 5805
$\AA$ features. The slightly higher spectral resolution of the INT (middle) spectrum
shows clearly  \ion{[N}{ii]} 6548, 6584 $\AA$ components contributing to H$\alpha$. 
There is no indication of a significant variation, from one spectrum to another, 
of the high-excitation lines, which could be related to variability of the exciting
source.  \\

When comparing with the latest available spectrum in the litterature, obtained
in 1990 (see Ringwald et al., \cite{Ring96}), we can note that the high 
excitation lines (HeII, CIV) are still  strong, but that the [OIII] 
$\lambda\lambda$ 4959, 5007 $\AA$ lines, present in addition to the classical
accretion disk lines, are much weaker now, as expected  
because of the expansion of the nebula.  In practice, we 
detect only one, faint  [OIII] component, blueshifted by 500 km s$^{-1}$, which
may correspond to one of the nebular spots detected by Harman \& O'Brien
(\cite{Harm03}). \\
 
The equivalent width of H$\alpha$ has also been measured on our
low-dispersion spectra: the comparison of the long-slit,
integrated values with  the high spectral resolution values found with the 3"
circular aperture, can give an indication of the contribution of the nebula
to the measurements. When integrating all emission along the slit
(extension about 13") in our December 2008 spectrum, for instance, we find a
total equivalent width of  H$\alpha$ of -20 $\AA$, which is larger than any
value found on high-dispersion spectra at the same epoch. 
But if we restrict the integration in the long slit spectrum 
to the central 3" only, we find -16 $\AA$ only, in good agreement with the
high-resolution value.  There is therefore at most a 20\% contribution of
something else to the central value extending to large radii from the
central object. 

The INT spectrum, obtained with a 1".5 slit in good seeing conditions
($\sim 1".0$) similarly yields an equivalent width for the central core of the 
H$\alpha$ line of -13.4 $\AA$, in excellent  agreement with the high-resolution
spectra core values. We can thus safely consider that most of the H$\alpha$
emission comes from the very central parts, and that 
10\% is a conservative upper limit for the possible contribution of the
nebula (or the nebular blobs) to the  equivalent width of the central
component of H$\alpha$ in the high-resolution spectra. \\

\section{Discussion}

The first point to note is that the real spectroscopic phases of
\object{HR Del} are uncertain. K\"urster and Barwig (\cite{Kurs88})
estimate an
error of 5 10$^{-6}$ days for a spectroscopic period of 0.214165 days.
Their observations were obtained in 1978-1980, of the order of 9000
days before our
observations, so the phase uncertainty since then could be of the
order of one orbital period.  We can thus only be reasonably
certain about the relative spectroscopic phases at each epoch of our
observations, but we can seek to obtain information about
the real phases using another approach.
 It should be noted that K\"urster and Barwig (\cite{Kurs88})
used a convention for spectroscopic binaries, with zero phase
at maximum radial velocity, which is different from the one generally used for cataclysmic binaries. 
The photometric data used the convention
for eclipsing variables, with zero phase at the minimum of the light
curve (as in Fig.~\ref{li-curva}, the zero point being given in the
figure caption). In fact, by comparing their zero point with our photometric, zero point, 
we find that the K\"urster and Barwig
phase (NOT the true spectroscopic phase) corresponding to our photometric zero point, zero phase 
(as in our  Fig.~\ref{li-curva})  is 0.42. \\ 

A particularly striking feature of our spectra is that the H$\alpha$ profile
seems to be a sum of the already mentioned central and wing components.
It is hard to make a comparison with the lower resolution spectra of
K\"urster
and Barwig (\cite{Kurs88}) with resolutions of 1 and 2 $\AA$, taken from
1978
to 1980, as those spectra had still serious interpretation problems due to
contamination by nebular emission. \\

We can compare the  radial velocities corresponding to the widths of the
different components of our high resolution H$\alpha$ profiles  with the
velocities of the emission produced by the expanding nebula. Harman
and O'Brien (\cite{Harm03}) find a maximum of about 600 kms$^{-1}$ for the
radial velocity  of major nebular  H$\alpha$ emission, such a low velocity
being expected for this slow nova.  Moraes and Diaz
(\cite{Mora09}) also find that the highest radial velocity of  H$\alpha$
emitting material is  associated with the polar caps, and has  a radial
velocity of 630 km s$^{-1}$.
If a nebular contribution were important, our H$\alpha$ profile
would be expected to show a component with about twice this width in
velocity units, yet the dominant central component is much narrower. 
The limits in radial velocity of the line centre and line wing
components are easily measurable in our spectra, as seen  for instance
in Fig.~\ref{high}:  the violet edges are at about -250 and -650
kms$^{-1}$ for the central component and the blue wing
respectively, and similar values are found on the red side (although 
there they could be slightly modified due to a possible  contamination
by \ion{[N}{ii]}). We thus see that a major nebular contribution is
implausible for the
central component, but cannot  be easily excluded in the wings. \\
Weak 6548.1 and 6583.6 \AA\ \ion{[N}{ii]} lines
are also present in the nebular emission (Harman and O'Brien
\cite{Harm03}, Moraes and Diaz \cite{Mora09}).
There is no evidence in our spectra (e.g. Fig.~\ref{high}) for a central
peak due to these lines, that is neither for the fainter,
6548.1 \AA\ line,  nor for the line
which is three times stronger at 6583.5 $\AA$.
The nebular \ion{[N}{ii]} contribution can thus be deemed undetectible.
Separation of the nebular component may be easier in
the future with space resolved spectroscopic observations of the
centre of the \object{HR Del} image.
\\

With an inclination of 42$^o$, a white dwarf mass of 0.65$M_{\odot}$ and a
radial velocity of 250 km $s^{-1}$ (measured from our spectra), the radius of the
outermost circular orbit of emission for the central component of the
profile 
would be at 6.2 $10^{10}$ cm or 0.89 $R_{\odot}$ from the centre of the
white dwarf
stellar component. This is significantly larger than the radius of the
photo-ionising disk found by  Moraes and Diaz (\cite{Mora09}). The reason
for this difference is not clear at present; parts of the disk at
intermediate radii
may not produce much ionising radiation and/or may only contribute to
the profile wings.
It must be noted that the radius of the emitting region appears somewhat
too large
when compared with the equatorial radius for an accretion disk around
the white
dwarf according to the two possible system parameters of K\"urster and
Barwig
(\cite{Kurs88}), but the distance from the centre of the white dwarf to the
inner Lagrangian point is close to the calculated value of the  outer
radius of the disk,
using expressions on page 33 of Warner (\cite{Warn95}). The simplest
explanation of these results is that some radiation is emitted by regions of
higher radial velocity at the outer edge of the disk than that directly
deduced from the observations, but that it is occulted because of
a particular geometry of the line emitting region. \\

High velocity components have been observed in \object{HR Del} in the
past, in particular in the high velocity \ion{C}{iv} resonance
line P Cygni absorption
component in the ultraviolet, which had an edge radial velocity of
-5000 km s$^{-1}$ (Selvelli and Friedjung, \cite{Selv03}).
Such a component could have been produced in a high velocity jet.
No evidence  for it is seen today in our
H$\alpha$ profiles, nor in older optical profiles such as that
of the 1990 spectrum (Ringwald et al. \cite{Ring96}).
One can thus wonder to which extent the ultraviolet spectrum of HR Del
has changed
between the lifetime of IUE and the epoch of our optical spectra, but 
no UV spectra
could be obtained since the end of IUE in 1996. The only
recent, UV data
available are two photometric points from the Galex all-sky survey,
which give
the following AB magnitudes (observations of September 2006): 12.78 in
the Near-UV
(central wavelength 2315 $\AA$) and 12.96 in the Far-UV ($\lambda$~
1539 $\AA$). 
If taken at face value, these magnitudes would indicate a small fading
in the near-UV,
and a stronger one (about a factor of two) in the Far-UV with respect to
the average IUE
spectrum from Selvelli and Friedjung (\cite{Selv03}). As however
\object{HR Del} is
slightly brighter than the saturation limit of the Galex detectors,
those magnitudes have to be
corrected for dead-time effects. When this is done according to the
precepts described in
Morrissey et al. (\cite{Morr07}), the corrected AB magnitudes become
12.36 and 12.38 in the NUV
and FUV respectively. 

\begin{figure}
\centering
 \includegraphics[ width=\linewidth ]{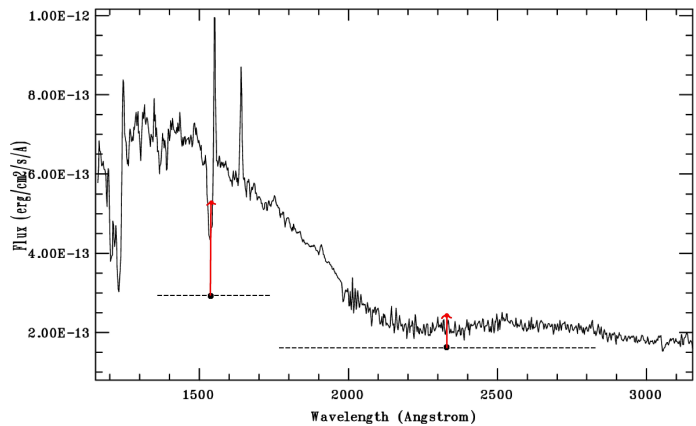}
\caption{The two Galex photometric points (lower, black points)
superposed on the IUE average spectrum from Selvelli and Friedjung
(2003). The horizontal dashed lines represent the bandpass of the two
(NUV and FUV) filters. The upper,
red points, linked to the corresponding observed (lower) values,
represent the real, estimated flux after correction for
deadtime effects in the detectors.  }
\label{UV}
\end{figure}

This means that the NUV magnitude is still about the same as in 1988,
within the uncertainties of the correction, and that in the FUV the
fading has
not exceeded 30 $\%$ since 1988, which is quite remarkable.
This is illustrated in Fig.\ref{UV}, which is Fig. 1 of Selvelli and
Friedjung
(\cite{Selv03}), where we have plotted the Galex values over the spectrum
(uncorrected for reddening). Of course, we still need a good UV spectrum
to assess what are  the real line profiles today. \\

We can note, in connection with non-orbital variations of H$\alpha$, that
Selvelli and Friedjung (\cite{Selv03}) suspected quite rapid non-orbital
variations of the \ion{C}{iv} and \ion{N}{v} ultraviolet resonance line
absorption in \object{HR Del}, similar to the rapid variations of
\ion{C}{iv}
seen for another old nova (\object{V603 Aql} in Friedjung et al.
(\cite{Fried97}) and in Prinja et al. (\cite{Prin00})). Such variations
of the
ultraviolet resonance absorption components of P Cygni profiles may be
expected
for the winds of accretion disks of cataclysmic binaries and have  been
observed  (Froning \cite{Fron05}, Proga \cite{Prog05}). \\

The "S wave type" pattern of H$\alpha$, displayed in
Figs.~\ref{SwaveB} to \ref{Swave}, is quite clear on the blue side, but somewhat less so 
on the red side of the profiles. It does however not appear to be a classical
cataclysmic binary S wave. However, it can, as it moves over the
central component of the line profile like the classical wave, be
understood as being near the outer edge of a rotating disk in the
neighborhood of a fixed point in the rotating frame of the binary.
The profile shows a maximum assymmetry towards the violet around
K\"urster and Barwig phase 0.75 or at photometric phase 0.33, that
is only  0.08 of an orbital period after photometric phase
0.25. If we interpret the photometic variations as due to varying
visibility of the regions of the surface of the mass loser,
heated by radiation from the disk and white dwarf (see below),
the photometric zero phase is when the mass loser is nearest
the observer in inferior conjunction. The phase of maximum
assymetry of the profile around photometric phase 0.33, is then not
far from quadrature when the spot is near a line perpendicular to the
line of sight. An accretion disk should rotate in the same direction
as the binary (conservation of angular momentum), so the H$\alpha$
bright spot should be {\it on the side of the disk opposite to the mass
loser}. Such a situation is not classical. That could possibly be
related to the large size of the accretion disk found above.\\

We can emphasize however that the stability of the S wave over the six
years of our spectroscopic observations (of the order of 10,000 orbital
cycles) indicates that it moves with a period which is  quite close to
the orbital period of K\"urster and
Barwig.  Indeed, using  their estimated error for the period, we calculate that the 
 phase shift over the time of our spectroscopic
observations would  be of the order of 0.23. Positive and negative
superhump periods, which can be permanent, would on the other hand
differ from the orbital period by a factor of the order of a few
percent when the orbital period is
that of \object{HRDel} (Olech et al. \cite{Olec09}). \\

The variations of the V magnitude cannot be due to the aspect changes of
the distorted Roch lobe filling cool component in a normal state. The visual
magnitude was around 12.5 during our observations, which gives an absolute
magnitude around 2.35, using a distance of 850 pc and a visual extinction
of $Av = 0.5$ (Selvelli and Friedjung, \cite{Selv03}). An ordinary
star  not heated by its companion, with a mass of  0.53 solar masses
(mass of the companion, according to Selvelli and Friedjung,
\cite{Selv03}) should 
have an absolute magnitude of the order of 8.5 (assuming the star is on
the main sequence).
The contribution of an unheated
cool component to the V magnitude would thus be of the order of 1/300,
and variations
in its visibility could not produce the observed photometric changes in
V with
a full amplitude of 0.08 magnitudes. It would moreover be hard to
produce the observed symmetrical light curve, by supposing extra light
produced by an assymmetrical bright spot.\\
 
If the ellipticity (or, more precisely, the shape of the Roche lobe) of
the cool component produced the photometric variations, this component would
then be strongly heated by the hot component, consisting of a white
dwarf plus an accretion disk. As we have seen this interpretation poses
a problem in understanding the  phasing of the H$\alpha$ bright spot.
Such an interpretation disagrees with the relative
phases of the S wave and the photometric variations unless the S wave
is not produced by a classical bright spot. A more detailed discussion
of this aspect is beyond the scope of this paper and will be
given in a forthcoming one (Voloshina et al., in preparation). \\

\begin{figure}
\centering
 \includegraphics[bb=16 16 237 312, width=7.5cm, angle=0,
clip]{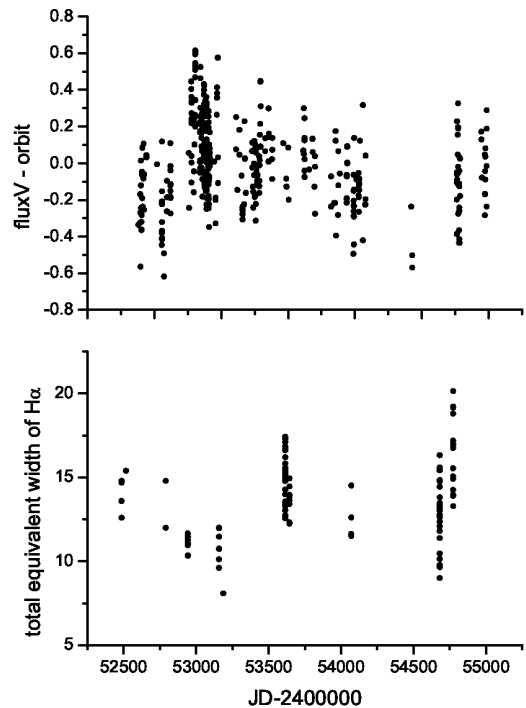}
\caption{Combined longer timescale photometric and spectroscopic
behaviour. The upper part shows the V magnitude (corrected for orbital variations) versus Julian date, 
and the lower part the total equivalent width of  H$\alpha$.  }
\label{PhotSpec}
\end{figure}

The large ``activity'' of HR Del found by Selvelli and Friedjung
(\cite{Selv03}), and confirmed by the high luminosty determined by
Moraes and 
Diaz (\cite{Mora09}), still appears to be unusual, whatever
explanation is
given for it. The high luminosity leads to heating of the mass  loser. \\

Long time scale variations are apparent both in the total magnitude of the object, and in the equivalent width 
of H$\alpha$ (see Fig.~\ref{PhotSpec}), but the lack of obvious relation between the two does not give us any clues 
as to their origin. 
Those long time scale variations  cannot be similar to those of
dwarf novae, which, when not in outburst, are fainter than recent old
novae and not
brighter like \object{HR Del}. The behaviour of dwarf novae is
understood as due
to a lower accretion rate of the white dwarf component outside outburst,
compared with
that of old novae. Present theories of dwarf nova outbursts explain them by
accretion disk instabilities, which can occur at low accretion rates. Small
outbursts, somewhat similar to those of dwarf novae, have been observed
for the
classical novae \object{GK Per}  and \object{V446 Her}  some decades
after the outburst; this is
understood as due to a drop of the accretion rate at such times, associated
with the appearence of instability. The situation is less clear for
the low
amplitude (less than 1 magnitude) ``stunted'' outbursts seen for other old
novae and novalike systems (Honeycutt et al. \cite{Hone98}).

 Solar type cycles of the mass loser, leading to variations in the
accretion rate over time scales of the order of a decade, have also been
suggested as an explanation of long term changes in the luminosity of
cataclysmic variables (Bianchini, \cite{Bian87} and Warner,
\cite{Warn88}). It
remains to be seen whether this is possible in the presence of strong
irradiation of the mass loser, if one wishes to explain the long timescale
variations of \object{HR Del} in this way. \\

\section{Conclusions}

 We have performed photometric, and high resolution H$\alpha$ spectroscopic
observations of HR Del from 2002 to 2008. The photometry shows
variations depending on orbital phase, plus apparently random variations
from
night to night and perhaps variations with a longer time scale of
several years. The
orbital phase dependent variations may be interpretable by irradiation
of the
mass loser by a still very bright expanded white dwarf and its accretion
disk.
It is not clear at present whether the long time scale variations are
interpretable in terms of solar type cycles of the mass loser. \\
      Ultraviolet, broad-band photometric data from the GALEX archive
suggest only a small fading since the epochs of the IUE, but no recent
spectra are available to estimate the present activity from the UV
spectral lines. \\

  The H$\alpha$ profile shows both a narrow and a wide component. The 
former appears to be due to emission by an accretion disk, while the latter
probably contains
a non-negligible contribution due to the expanding nebula. The central
component does not show  a clear dependance of
equivalent width with  orbital phase. Its profile however shows  a clear
orbital
variation, which looks similar to the S wave generally seen in spectra of
cataclysmic binaries, but which has the wrong phase dependence to be
produced
by a classical bright spot. The stability of the S wave with respect to
the orbital period over 8 years indicates no connection with the period of
any positive or negative superhump. The exact nature of our "S wave"
still remains
to be elucidated.\\
     In view of the exceptional brightness and activity of this old nova
about 35 years after its eruption, continuing
low level thermo-nuclear combustion in the envelope of a low mass white
dwarf may
be considered a possible explanation. The evolution of this interesting
stellar
system should therefore be closely followed over a
large wavelength range in the future. \\

\begin{acknowledgements}
I. Voloshina acknowledges support from the CNRS-EARA which sponsored her visits to the IAP, where part 
of this work was completed. M. Dennefeld and M. Friedjung thank D. Briot who kindly cooperated in obtaining  
 a long series of spectra in August 2008, and  
the OHP night assistants (most notably R. Giraud and JC Mevolhon) for 
their valuable contribution to the service observations. 
The authors are gratefull to an anonymous referee for usefull suggestions. \\
Galex is a NASA Small Explorer, developed in cooperation with the French space agency CNES,  and 
the Korean Ministry of Science and Technology.
\end{acknowledgements}


\begin{appendix}

A journal of the CCD photometric observations is presented in Table~\ref{ObsPhot}. 
Observing data for individual 
 spectra are presented in Table~\ref{ObsSpec} and ~\ref{ObsSpec2}. All these tables are available 
only in electronic form. 

\vfill
\newpage

\begin{table}[htb]
\caption{Journal of the CCD observations}
\label{ObsPhot}
\centering
\begin{tabular}{rcccr}
\hline
Date	&set start-end&bands&exposure& N \\
		&24000000+    &    &        & \\
\hline
25.08.2005 &53608.511-53608.586&V&30s&135 \\
26.08.2005 &53609.558-53609.582&V&40s& 39 \\
27.08.2005 &53610.554-53610.578&V&50s& 35 \\
30.08.2005 &53613.541-53613.571&V&60s& 32 \\
31.08.2005 &53614.341-53614.371&V&30s& 61 \\
01.10.2005 &53645.259-53645.325&V&40s& 64 \\
03.10.2005 &53647.234-53647.280&V&40s& 80 \\
13.10.2005 &53657.304-53657.342&V&30s& 36 \\
26.07.2006 &53943.527-53943.594&V&40s&121 \\
27.07.2006 &53944.542-53944.592&V&40s& 84 \\
02.08.2006 &53950.538-53950.598&V&40s& 93 \\
03.08.2006 &53951.557-53951.604&V&40s& 73 \\
04.08.2006 &53952.527-53952.602&V&40s&121 \\
06.08.2006 &53954.508-53954.584&V&40s&113 \\
26.09.2006 &54005.224-54005.494&V&30s&615\\
27.09.2006 &54006.216-54006.463&V&30s&561\\
28.09.2006 &54007.417-54007.473&V&30s&130 \\
14.10.2006 &54023.172-54023.408&V&40s&527 \\
18.10.2006 &54027.313-54027.430&V&40s&193 \\
19.10.2006 &54028.314-54028.428&V&40s&192 \\
20.10.2006 &54029.320-54029.406&V&40s&182 \\ 
21.10.2006 &54030.319-54030.411&V&40s&161 \\
22.10.2008 &54762.191-54762.402&V,R&60s,60s&116 \\
23.10.2008 &54763.182-54763.371&V,R&60s,60s& 84 \\ 
03.11.2008 &54774.158-54774.351&V,R&80s,80s&104 \\
04.11.2008 &54775.194-54775.267&V,R&80s,80s& 33 \\
07.11.2008 &54778.149-54778.338&V,R&60s,80s&116 \\
10.11.2008 &54781.216-54781.380&R&200s& 50   \\
11.11.2008 &54782.166-54782.374&R&240s&83   \\
12.11.2008 &54783.155-54783.360&R&200s&82 \\
14.11.2008 &54785.151-54785.282&R&240s&41 \\
11.10.2009 &55116.221-55116.440&V&150s&104 \\
12.10.2009 &55117.185-55117.433&V&150s&117 \\
13.10.2009 &55118.186-55118.318&V&150s &64 \\
\hline
\end{tabular}
\end{table}

\vfill
\newpage

\begin{table}
\caption{Details of spectroscopic observations}
\label{ObsSpec}
\centering
\begin{tabular}{c c c c c}
\hline\hline
Date		&Grating&UT time of	&Orbital 	&Equivalent \\
		&(l/mm)	&exposure	&phases		& Width of H$\alpha$\\
		&	&		&		&Total (Central) \\
\hline
8/8/2002	&1200	&00h50		&0.698		&-13.6 (11.6) \\
		&	&01h24		&0.808		&-12.6 (11.3) \\
		&	&02h05		&0.941		&-14.7 (13.1) \\
		&	&02h38		&0.048		&-14.8 (13.2) \\
24/9/2002	&1200	&23h56		&0.649		&-15.4 (9.5) \\
13/6/2003	&1200	&01h07		&0.566		&-12.0 (11.2) \\
		&	&02h02		&0.744		&-14.8 (13.4) \\
1/11/2003	&1200	&17h53		&0.198		&-11.7 (9.3) \\
		&	&18h25		&0.302		&-11.0 (8.8) \\
		&	&18h58		&0.409		&-10.3 (7.3) \\
		&	&19h40		&0.545		&-10.3 (7.6) \\
		&	&20h15		&0.659		&-11.0 (7.6) \\
		&	&20h55		&0.789		&-11.1 (7.3) \\
		&	&21h27		&0.892		&-11.3 (6.8) \\
		&	&22h00		&0.999		&-11.7 (8.7) \\
2/11/2003	&1200	&18h20		&0.955		&-11.5 (8.2) \\
		&	&18h54		&0.066		&-11.3 (8.8) \\
		&	&19h40		&0.215		&-10.9 (7.6) \\
		&	&20h16		&0.331		&-11.1 (7.7) \\
6/6/2004	&1200	&01h12		&0.859		&-9.6 (7.6) \\
		&	&01h42		&0.957		&-10.1 (7.8) \\
		&	&02h12		&0.054		&-10.8 (7.5) \\
29/6/2004	&1200	&00h37		&0.140		&-12.0 (8.8) \\
		&	&01h07		&0.237		&-12.0 (8.9) \\
		&	&01h40		&0.344		&-11.5 (8.2) \\
		&	&02h11		&0.445		&-10.7 (7.4) \\
25/7/2004	&1200	&02h04		&0.824		&-8.1 (5.1) \\
6/10/2004	&1200	&19h10		&0.009		&-12.9 (11.3) \\ 
		&	&20h00		&0.171		&-13.9 (10.5) \\ 
		&	&20h34		&0.282		&-14.9 (11.2) \\ 
		&	&21h09		&0.395		&-14.0 (10.7) \\  
10/9/2005	&1800	&22h07		&0.475		&-12.7 (12.1) \\
		&	&22h44		&0.595		&-12.6 (11.8) \\
11/9/2005	&1800	&20h19		&0.794		&-14.3 (9.9) \\
		&	&21h20		&0.992		&-13.6 (9.9) \\
		&	&22h20		&0.186		&-15.2 (10.3) \\
		&	&23h21		&0.384		&-13.0 (9.0) \\
12/9/2005	&1800	&00h33		&0.617		&-12.8 (8.8) \\
13/9/2005	&600	&20h16		&0.123		&-15.5 (12.2) \\
		&	&20h47		&0.223		&-15.8 (12.0) \\
		&	&21h19		&0.327		&-14.0 (10.2) \\
		&	&21h47		&0.418		&-14.8 (10.1) \\
		&	&22h17		&0.515		&-14.3 (10.1) \\
		&	&22h48		&0.616		&-15.6 (11.2) \\
		&	&23h18		&0.713		&-15.4 (11.3) \\
		&	&23h48		&0.810		&-13.3 (9.2) \\
14/9/2005	&600	&00h18		&0.907		&-17.4 (12.6) \\
26/9/2005	&600	&20h32		&0.875		&-15.3 (10.9.6) \\
		&	&21h06		&0.986		&-16.8 (12.8) \\
		&	&21h36		&0.083		&-16.6 (12.9) \\
		&	&22h06		&0.180		&-17.1 (12.9) \\
		&	&22h37		&0.281		&-16.2 (12.6) \\
		&	&23h08		&0.381		&-13.5 (9.4) \\ 
27/9/2005	&600	&21h20		&0.700		&-14.9 (10.8) \\
		&	&21h51		&0.801		&-15.0 (11.6) \\
		&	&22h23		&0.905		&-16.7 (12.8) \\
		&	&22h54		&0.005		&-17.3 (12.1) \\
\hline
\end{tabular}
\end{table}

\begin{table}
\caption{Details of spectroscopic observations, continued}
\label{ObsSpec2}
\centering
\begin{tabular}{c c c c c}
\hline\hline
24/10/2005	&600	&18h24		&0.201		&-13.4(10.4) \\
		&	&18h55		&0.301		&-14.9(11.4) \\
		&	&19h26		&0.402		&-13.7(9.6) \\
26/10/2005	&600	&18h12		&0.500		&-12.3 (10.0) \\
		&	&18h42		&0.598		&-12.3 (9.8) \\
		&	&19h12		&0.695		&-13.4 (10.7) \\
		&	&19h42		&0.792		&-13.8 (11.3) \\
		&	&20h10		&0.883		&-14.5 (10.7) \\
		&	&20h44		&0.993		&-13.9 (9.3) \\ 
		&	&21h07		&0.068		&-14.0 (9.8) \\
14/12/2006	&1200	&17h48		&0.512		&-11.5 (9.2) \\
		&	&18h22		&0.622		&-12.6 (9.6) \\    
		&	&19h33		&0.852		&-11.6 (8.6) \\
		&	&20h10		&0.972		&-14.5 (9.9) \\
26/8/2008	&1200	&21h13		&0.810		&-9.0 (6.9) \\
		&	&21h47		&0.920		&-10.1 (8.0) \\
		&	&22h18		&0.020		&-11.4 (8.2) \\
		&	&23h00		&0.157		&-11.0 (8.7) \\
		&	&23h31		&0.257		&-10.5 (8.0) \\
27/8/2008	&1200	&00h03		&0.361		&-9.7 (7.1) \\
		&	&02h08		&0.766		&-9.8 (5.4) \\
		&	&21h22		&0.508		&-13.2 (11.2) \\
		&	&21h55		&0.615		&-13.5 (11.8) \\
		&	&22h36		&0.748		&-13.4 (11.4) \\
		&	&23h09		&0.855		&-14.8 (13.1) \\
		&	&23h51		&0.991		&-15.6 (12.7) \\
28/8/2008	&1200	&00h22		&0.092		&-14.8 (12.3) \\
		&	&00h59		&0.212		&-12.6 (9.8) \\
		&	&01h31		&0.316		&-13.8 (11.7) \\
		&	&02h03		&0.419		&-15.4 (11.1) \\
		&	&02h41		&0.543		&-13.3 (10.5) \\
		&	&21h51		&0.272		&-14.8 (11.0) \\
		&	&22h25		&0.382		&-12.8 (10.3) \\
		&	&22h56		&0.482		&-12.1 (9.6) \\
		&	&23h35		&0.609		&-13.0 (10.0) \\
29/8/2008	&1200	&00h06		&0.709		&-12.4 (10.1) \\
		&	&00h36		&0.807		&-13.2 (10.6) \\
		&	&01h14		&0.930		&-14.4 (10.8) \\
		&	&01h45		&0.030		&-16.3 (10.9) \\
20/11/2008	&1200	&17h57		&0.734		&-13.9 (10.5) \\
		&	&18h28		&0.834		&-13.3 (9.9) \\
		&	&19h00		&0.938		&-16.9 (12.5) \\
		&	&19h30		&0.035		&-19.2 (14.2) \\
		&	&20h00		&0.133		&-18.8 (13.4) \\
		&	&20h30		&0.230		&-20.2 (14.0) \\
		&	&21h04		&0.340		&-19.2 (12.3) \\
		&	&21h35		&0.441		&-16.9 (11.4) \\
		&	&22h05		&0.538		&-14.3 (9.1) \\
21/11/2008      &1200   &18h48          &0.568          &-15.1 (10.3) \\
                &       &19h20          &0.672          &-14.0 (10.6) \\
                &       &19h50          &0.769          &-15.6 (10.8) \\
                &       &20h21          &0.870          &-17.2 (11.7) \\
                &       &20h53          &0.974          &-17.0 (11.9) \\
                &       &21h24          &0.074          &-16.8 (10.9) \\
                &       &21h58          &0.184          &-14.9 (9.0) \\
\hline
\end{tabular}
\end{table}

\end{appendix}

\end{document}